\documentclass[%
 aip,
 jcp,
 amsmath,amssymb,
 reprint,%
longbibliography,
]{revtex4-1}

\usepackage{graphicx}
\usepackage{dcolumn}
\usepackage{bm}
\usepackage[mathlines]{lineno}

\usepackage[utf8]{inputenc}
\usepackage[T1]{fontenc}
\usepackage{mathptmx}
\usepackage{etoolbox}
\makeatletter
\def\@email#1#2{%
 \endgroup
 \patchcmd{\titleblock@produce}
  {\frontmatter@RRAPformat}
  {\frontmatter@RRAPformat{\produce@RRAP{*#1\href{mailto:#2}{#2}}}\frontmatter@RRAPformat}
  {}{}
}%
\usepackage{hyperref}
\makeatother
\begin{document}

\preprint{AIP/123-QED}

\title{Two-dimensional partitioned square ice confined in graphene/graphite nanocapillaries}
\author{Zhen Zeng}
\affiliation{State Key Laboratory of Engines, Tianjin University, Tianjin, 300072, China.}
\author{Tianyou Wang}
\affiliation{State Key Laboratory of Engines, Tianjin University, Tianjin, 300072, China.}
\email{wangtianyou@tju.edu.cn}
\author{Rui Chen}
\affiliation{Department of Aeronautical and Automotive Engineering, Loughborough University, Loughborough LE11 3TU, United Kingdom.}
\author{Mengshan Suo}
\affiliation{State Key Laboratory of Engines, Tianjin University, Tianjin, 300072, China.}
\author{Kai Sun}
\affiliation{State Key Laboratory of Engines, Tianjin University, Tianjin, 300072, China.}
\author{Panagiotis E. Theodorakis}
\affiliation{Institute of Physics, Polish Academy of Sciences, Al. Lotnik\'{o}w 32/46, 02-668 Warsaw, Poland.}
\author{Zhizhao Che}
\email{chezhizhao@tju.edu.cn}
\affiliation{State Key Laboratory of Engines, Tianjin University, Tianjin, 300072, China.}

\date{\today}

\begin{abstract}
As one of the most fascinating confined water/ice phenomena, two-dimensional square ice has been extensively studied and experimentally confirmed in recent years. Apart from the unidirectional homogeneous square icing patterns considered in previous studies, the multidirectional partitioned square icing patterns are discovered in this study and characterized by molecular dynamics (MD) simulations. Square icing parameters are proposed to quantitatively distinguish the partitioned patterns from the homogeneous patterns and the liquid water. The number of graphene monolayers $n$ is varied in this study, and the results show that it is more energetically favorable to form partitioned square icing patterns when the water molecules are confined between graphite sheets ($n \geq 2$) compared to graphene ($n = 1$). This phenomenon is insensitive to $n$ as long as $n \geq 2$, because of the short-range nature of the interaction between water molecules and the carbon substrate. Moreover, it is energetically unfavorable to form partitioned square icing patterns for a single layer of water molecules even for $n \geq 2$, verifying that the interaction between layers of water molecules is another dominant factor in the formation of partitioned structures. The conversion from partitioned structure to homogenous square patterns is investigated by changing the pressure and the temperature. Based on the comprehensive MD simulations, this study unveils the formation mechanism of the partitioned square icing patterns.
\end{abstract}

\maketitle

\section{INTRODUCTION}\label{sec:sec1}
The phase behavior of water is a topic of perpetual interest not only because of its ubiquitousness but also owing to the important implications for biological, environmental, geological, and physical processes.\cite{Mishima1998SupercooledGlassyWater} There are at least 18 bulk crystalline ice phases from experimental verifications and several more from theoretical predictions.\cite{Debenedetti2003SupercooledGlassyWater, Huang2016UnderNegativePressure, Poole1992MetastableWater, Sciortino1995CrystalStabilityLimits, Zhao2014TwoDimensionalAmorphousIce, Zheng1991HomogeneousNucleationLimit} However, less focused but undoubtedly omnipresent is the confined water/ice, which is less obvious but all-important for its unique properties\cite{Compton2012MechanicalPropertiesofGraphene, Neek2016ViscosityofNanoconfinedWater} and many technological applications.

In nano-scale confinements, many new phases of crystalline ice have been reported. In particular, many low-dimensional crystalline ice phases have been recently revealed by experiments or simulations. For example, water confined in carbon nanotubes can form quasi-one-dimensional square,\cite{Koga2000IceNanotube} pentagonal,\cite{Bai2003PentagonandHexagonIce, Koga2001CarbonNanotubes, Koga2000IceNanotube} hexagonal,\cite{Bai2003PentagonandHexagonIce, Koga2001CarbonNanotubes, Koga2000IceNanotube} and heptagonal\cite{Byl2006UnusualHydrogenBonding, Koga2001CarbonNanotubes} single-walled ice nanotubes.\cite{Bai2003PentagonandHexagonIce, Byl2006UnusualHydrogenBonding, Koga2001CarbonNanotubes, Koga2000IceNanotube} Under higher hydrostatic pressure, water molecules can form high-density multi-walled helical ice nanotube.\cite{Bai2006MultiwalledIceHelixes} When confined between two parallel walls, water can form numerous two-dimensional monolayer ices, for instance, low-density octagonal monolayer ice and tetragonal monolayer ice,\cite{Bai2010MonolayerClathrate} mid-density hexagonal monolayer ice,\cite{Zhao2014FerroelectricHexagonal} high-density flat rhombic monolayer ice,\cite{Zhao2014FerroelectricHexagonal} high-density puckered rhombic monolayer ice,\cite{Bai2010MonolayerClathrate} and high-density square monolayer ice.\cite{Algara2015SquareIce} Multi-layer ice with the same patterns,\cite{Chen2017DoubleLayerIce, Corsetti2016HighDensityBilayer, Koga2005HydrophobicSurfaces, Koga2000AmorphousPhases} as well as new patterns,\cite{Algara2015SquareIce, Bai2012PolymorphismandPolyamorphism, Zhu2015CompressionLimit} has also been revealed, and the confined ice phases are stable only for a small range of plate separations.\cite{Zangi2004WaterConfinedReview} Some recent simulations shed light on low-dimensional ice phases by varying the temperature\cite{Zhu2016BucklingFailureofSquareIce, Zhu2017SuperheatingofMonolayerIce}, the lateral pressure,\cite{Zhu2016AbStacked, Zhu2016TrilayerIces, Zhu2016BucklingFailureofSquareIce} the separation between two graphene sheets,\cite{Zhu2015CompressionLimit, Zhu2016AbStacked} and the nanoscale confining surfaces,\cite{Qiu2015InhomogeneousNanoconfinement, Ruiz2018FlexibleConfiningSurfaces} and also by detailed analysis of the structural and dynamic characteristics.\cite{Zhu2017MonolayerSquareIce}

Experimental exploration of low-dimensional crystalline ice phases in nano-confined spaces has been carried out over the past decade. Yutaka et al.\cite{Maniwa2005PentagonaltoOctagonal} identified four distinct ordered polygonal ice-nanotube structures by systematic X-ray diffraction analysis, consistent with the results obtained by molecular dynamics (MD) simulations. Jinesh et al.\cite{Jinesh2008ExperimentalEvidence} provided experimental evidence for two-dimensional ice formation in nano-scale confinements at room temperature by a high-resolution friction force microscope. Algara-Siller et al.\cite{Algara2015SquareIce} reported the experimental evidence of monolayer and bilayer square ice between two graphene sheets at room temperature by transmission electron microscopy (TEM). When confined in nano-scale environments, water molecules form a square pattern which is qualitatively different from the conventional tetrahedral structure of hydrogen bonding between water molecules. Instead of temperature, the high van der Waals pressure becomes an important factor to induce the square icing structures.

Even though the existence of two-dimensional square ice has been proved and some important properties of it have been investigated, it should be noted that the square ice in graphene nanocapillaries has unidirectional homogeneous icing patterns in most of these previous studies. However, two-dimensional square ices with multiple partitions (i.e., the ice area is divided into regions in which the water molecules form the same square icing patterns but with different directions of hydrogen bonds) has never been analyzed. Even though the partitioned icing phenomenon is common in nature and the ice structure reported in the first experimental study of square ice does have a partitioned property.\cite{Algara2015SquareIce} In this study, we simulate the square ice between two parallel graphene-based sheets, and analyze the multidirectional, partitioned square icing patterns. From the perspective of the short-range nature of the interaction between water and carbon, the formation process of partitioned square structures, and the interaction between layers of water molecules, we unveil the formation mechanism of the partitioned square icing patterns. This study not only provides physical insights into the mechanism of the two-dimensional icing phenomena in graphene/graphite nanocapillaries, but also will be helpful for the practical applications in nanotribology, nanofluidic, and nanomaterials.

\section{COMPUTATIONAL METHODS}\label{sec:sec2}
\subsection{MD Simulation}
\begin{figure}
  \centering
  \includegraphics[width=0.85\columnwidth]{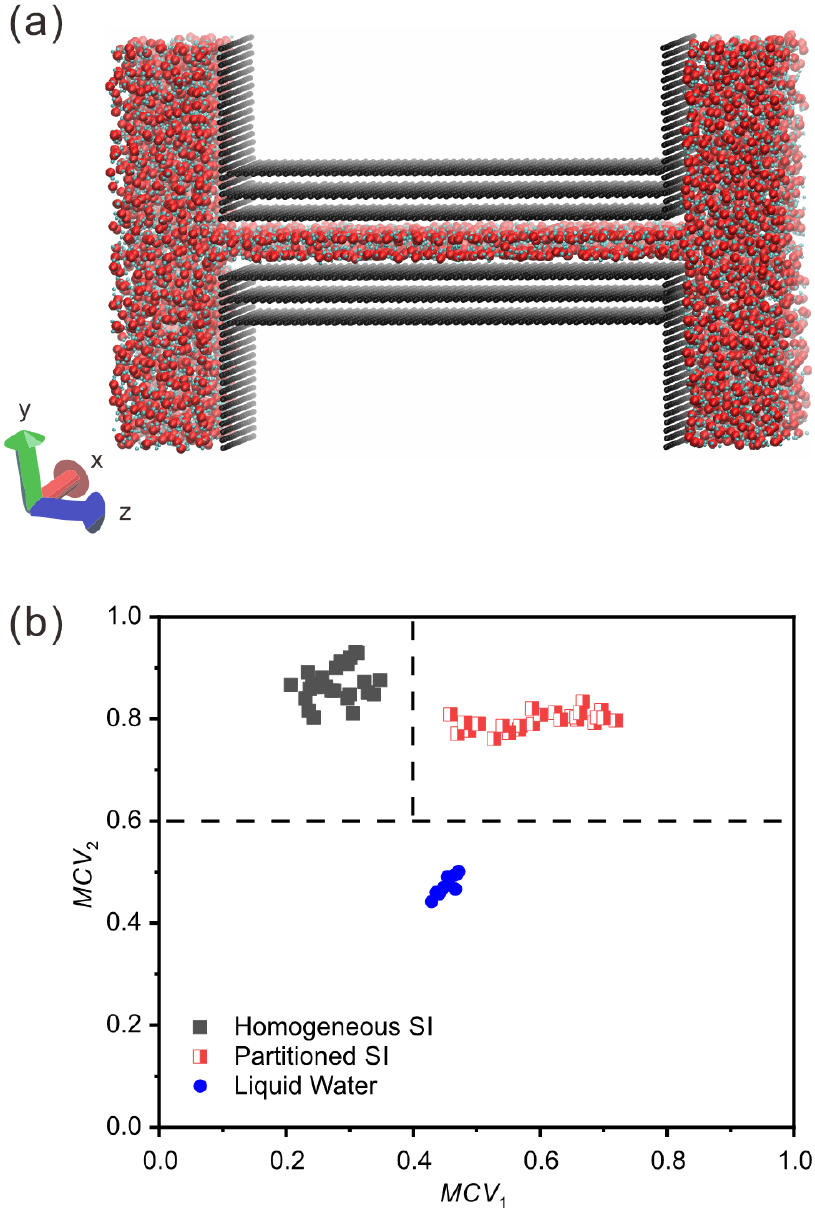}\\
  \caption{(a) Initial configuration of MD simulation. The red beads represent oxygen atoms, the cyan beads represent hydrogen atoms, and the black beads represent carbon atoms. (b) Square icing parameter map. $MCV_1$ and $MCV_2$ are square icing parameters to analyze the uniformity of the distribution of water molecules in spatial position and direction of the hydrogen-oxygen bonds. They are defined in Eq. (1) and (2) and able to identify the different configurations of water molecules. Each symbol represents a complete simulation case: the black solid squares represent the simulation cases with homogeneous square icing patterns, the half-solid-half-hollow red squares represent the simulation cases with partitioned square icing patterns, and the blue solid circles represent the cases with disordered liquid water.}\label{fig:fig01}
\end{figure}

The initial configuration of the MD simulation is shown in Figure \ref{fig:fig01}(a), and is similar to that used in Ref.\ \cite{Algara2015SquareIce}. A long capillary is formed by two parallel graphene-based sheets consisting of different numbers of graphene monolayers. We use the most common graphene stacking ABA in our multilayer simulation systems. (We have also tested the AAA stacking for comparison to find out how the stacking affects the ice patterns. The result shows that square ice is formed independently of whether the graphene stacking is AAA or ABA. See Figure S1 in Supplementary Material for details.) Two water reservoirs with each containing 2000 water molecules are connected by the capillary. The number of graphene monolayers on each side of the water molecules is varied between $1 \leq n \leq 6$. The length and the width (in $z$ and $x$ direction as shown in Figure \ref{fig:fig01}(a)) of the graphene capillary are fixed at 68 {\AA} and 56 {\AA}, respectively. The height $h$ of the graphene capillary (i.e., the distance between the two graphene sheets in $y$ direction) is chosen to be 6.5 {\AA} or 9.0 {\AA} to accommodate one or two layers of water molecules, respectively. The carbon atoms in the graphene sheets are fixed during simulations, as previous studies have shown that square ice is formed independently of whether the confinement is provided by rigid or flexible graphene sheets.\cite{Algara2015SquareIce, Ruiz2018FlexibleConfiningSurfaces}

MD simulations are performed in the isothermal, isobaric ensemble (NPT), in which the temperature (298 K) and the pressure are controlled by the Nos\'{e}-Hoover thermostat and barostat, respectively. Periodic boundary conditions are applied in the three directions of the simulation box. In the equilibration run, the lateral pressure $P$ is kept at 1 bar ($10^{-4}$ GPa) for the first 5 ns, during which water molecules fill in the graphene capillary. After that, $P$ is increased to 1.1 GPa (critical lateral pressure under which the square icing patterns can form, as shown in Figure S2 in Supplementary Material) and maintain for more than 25 ns, considering the extra lateral forces acting on water molecules in experiments.\cite{Algara2015SquareIce} The temperature is fixed at 298 K during the simulations. A time step of 1.0 fs is used for the velocity-Verlet integrator. Water molecules are modeled using the extended simple point charge (SPC/E) model, including the long-range Coulomb potential and the short-range Lennard-Jones potential between the interaction sites.\cite{Berendsen1987EffectivePairPotentials} The SPC/E water model has been proved to be accurate in the simulations of monolayer and bilayer square ice structure by comparing with experimental results.\cite{Algara2015SquareIce} The water-carbon interaction is modeled by a 12-6 Lennard-Jones potential between the carbon and the oxygen atoms. The distance and the energy parameters are $\sigma = 0.328$ nm and $\epsilon = 0.114$ Kcal/mol,\cite{Che2017SurfaceNanobubbles, Gordillo2000HydrogenBondStructure, Joly2011GiantLiquidSolidSlip, Werder2008WaterCarbonInteraction} respectively, and the distance between graphene monolayers is 3.4 {\AA} in graphite sheets. The long-range interactions are computed using the particle-particle particle-mesh (PPPM) algorithm with an accuracy of 10$^{-4}$. The simulations are carried out by using LAMMPS,\cite{Plimpton1995FastParallelAlgorithms} and the snapshots are rendered in VMD.\cite{Humphrey1996VMD}

\subsection{Square Icing Parameters}
To quantitatively determine and distinguish the partitioned square icing pattern from the homogeneous square icing pattern and the liquid water, square icing parameters, $MCV_1$ and $MCV_2$, are proposed in this study:
\begin{equation}\label{eq:eq01}
MC{{V}_{1}}=\frac{1}{M}\sum\limits_{m=1}^{M}{\frac{{{\left[ \frac{1}{L}{\sum\limits_{l=1}^{L}{\left( {{A}_{lm}}-{{{\bar{A}}}_{m}} \right)}^{2}} \right]}^{\frac{1}{2}}}}{{{{\bar{A}}}_{m}}}}
\end{equation}
\begin{equation}\label{eq:eq02}
MC{{V}_{2}}=\frac{1}{L}\sum\limits_{l=1}^{L}{\frac{{{\left[ \frac{1}{M}{\sum\limits_{m=1}^{M}{\left( {{A}_{lm}}-{{{\bar{A}}}_{l}} \right)}^{2}} \right]}^{\frac{1}{2}}}}{{{{\bar{A}}}_{l}}}}
\end{equation}
where $L$ and $M$ are two free positive integers, $l$ is an integer that runs from 1 to $L$, $m$ is an integer that runs from 1 to $M$, and $A_{lm}$ is the number of water molecules whose coordinate and orientation satisfy:
\begin{equation}\label{eq:eq03}
{{Z}_{\min }}+{\left( l-1 \right)\left( {{Z}_{\max }}-{{Z}_{\min }} \right)}/{L}\;\le z<{{Z}_{\min }}+{l\left( {{Z}_{\max }}-{{Z}_{\min }} \right)}/{L}
\end{equation}
\begin{equation}\label{eq:eq04}
{\left( m-1 \right)\pi }/{\left( 2M \right)}\;\le \theta <{m\pi }/{\left( 2M \right)}
\end{equation}
where $z$ is the coordinate of the water molecule in $z$ direction, $Z_\text{min}$ and $Z_\text{max}$ are the coordinate minimum and maximum of the graphene capillary in $z$ direction, respectively, and $\theta$ represents the direction of the hydrogen-oxygen bonds in the water molecule, which runs from $\theta = 0$ to $\theta = \pi/2$. The overbar in Eqs.\ (\ref{eq:eq01}) and (\ref{eq:eq02}) indicates the average, i.e.,
\begin{equation}\label{eq:eq05}
{{\bar{A}}_{m}}=\frac{1}{L}\sum\limits_{l=1}^{L}{{{A}_{lm}}}
\end{equation}
\begin{equation}\label{eq:eq06}
{{\bar{A}}_{l}}=\frac{1}{M}\sum\limits_{m=1}^{M}{{{A}_{lm}}}
\end{equation}

These square icing parameters analyze the uniformity of the distribution of water molecules in spatial position and direction of the hydrogen-oxygen bonds, so as to be able to identify the different configurations of water molecules. From the data of many MD simulations (all of our simulation results including, but are not limited to, the simulation cases presented in this article), a map of the square icing parameters is generated, as shown in Figure \ref{fig:fig01}(b), in which $L = 13$ and $M = 18$ are chosen based on trials. The partitioned square ice (partitioned SI) can be easily distinguished from the homogeneous square ice (homogeneous SI) and liquid water when we take $MCV_1 = 0.4$ and $MCV_2 = 0.6$ as thresholds.

\section{RESULTS AND DISCUSSION}
\subsection{Partitioned Square Icing Phenomenon}
\begin{figure}
  \centering
  \includegraphics[width=0.9\columnwidth]{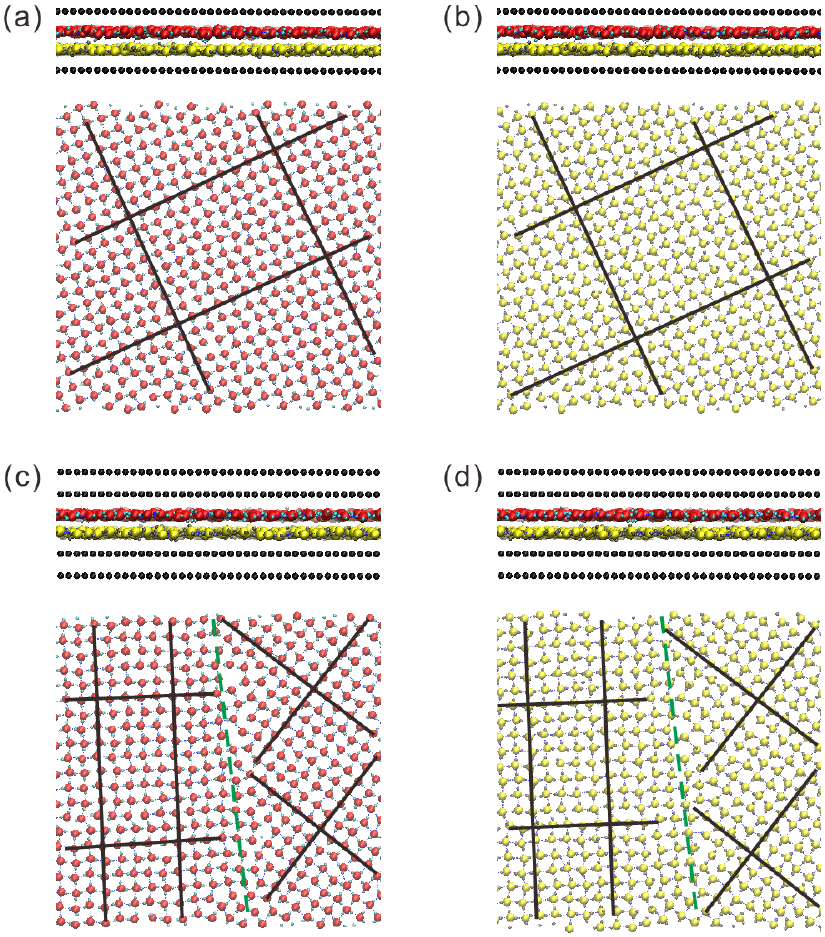}\\
  \caption{Side and top views of the simulation results. (a) Side and top views of water molecules in the top layer when $n = 1$. (b) Side and top views of water molecules in the bottom layer when $n = 1$. (c) Side and top views of water molecules in the top layer when $n = 2$. (d) Side and top views of water molecules in the bottom layer when $n = 2$. The red and the yellow beads represent oxygen atoms, the cyan and the gray beads represent hydrogen atoms, while the black beads represent carbon atoms. The blue dashed lines represent hydrogen bonds, the black solid line is a guide of the icing direction to the eye and the green dashed line is a guide to distinguish different regions.}\label{fig:fig02}
\end{figure}

\begin{figure*}
  \centering
  \includegraphics[width=1.4\columnwidth]{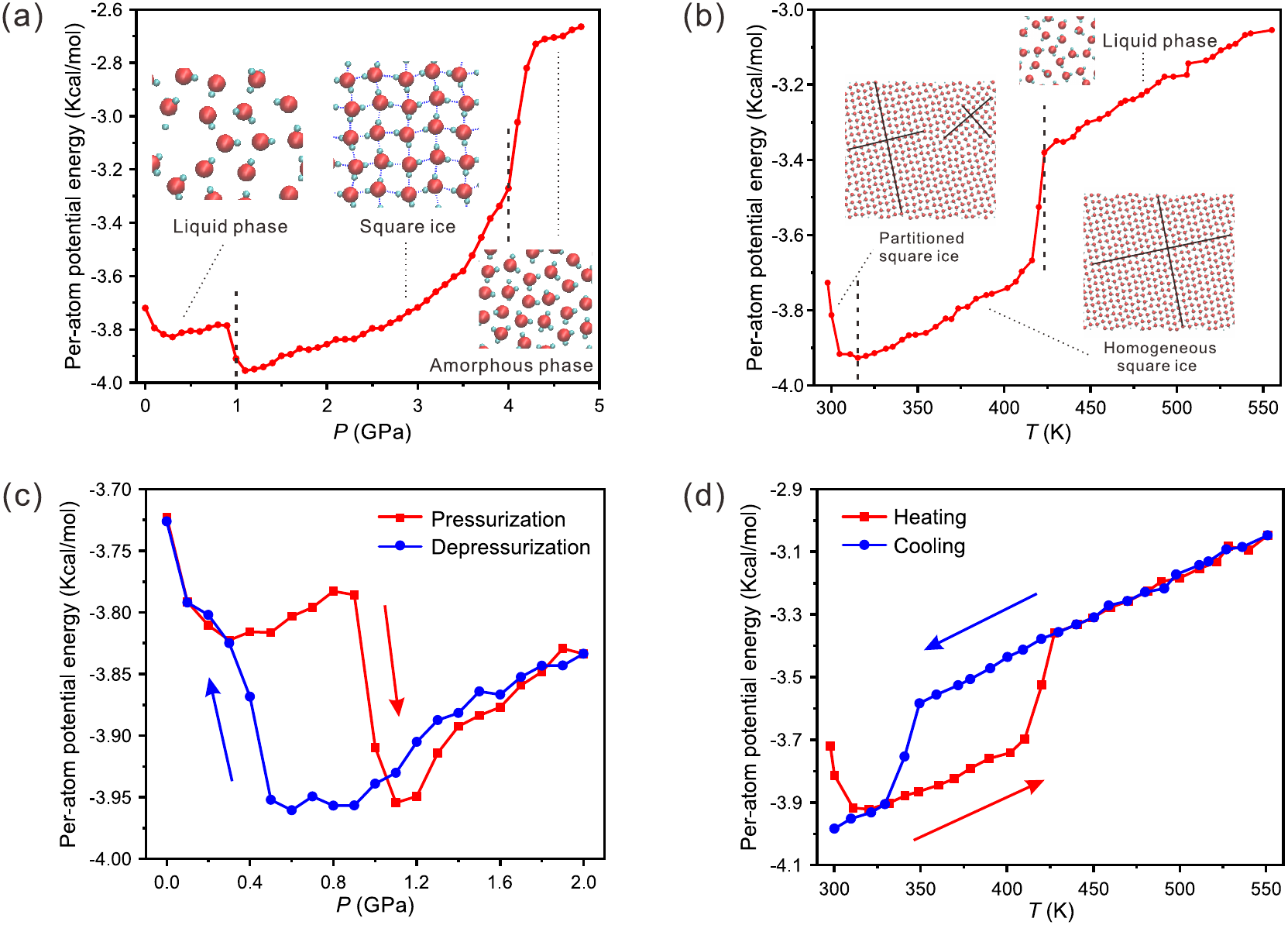}\\
  \caption{Variation of the potential energy of the confined water during the pressurization (or depressurization) and heating (or cooling) process. The red beads represent oxygen atoms, the cyan beads represent hydrogen atoms, the blue dashed lines represent hydrogen bonds and the black solid line is a guide of the icing direction to the eye. (a) Pressurization process; (b) Heating process; (c) Pressurization/depressurization process; (d) Heating/cooling process. Video clips for the processes corresponding to Figure \ref{fig:fig03}(a) and (b) are available in the Supplementary Material as Movies S2 and S3.}\label{fig:fig03}
\end{figure*}

We carried out a series of MD simulations of water confined between two parallel graphene-based sheets, which consisted of different numbers of graphene monolayers. When there is only one graphene monolayer in the sheet, the confined water molecules form homogeneous square icing patterns in tune with the results revealed in Ref.\cite{Algara2015SquareIce}. From the side and the top views of the simulation results, shown in Figure \ref{fig:fig02}(a) and (b), we can see the square icing patterns with hydrogen bonds in two perpendicular directions, and the water molecules in the top and the bottom layers form exactly the same structure. When we add another graphene monolayer, i.e., changing the number of graphene monolayers to two, the water molecules between them exhibit partitioned square icing patterns, as shown in Figure \ref{fig:fig02}(c) and (d). We can observe two distinct regions in which the water molecules both formed obvious square icing patterns but with different directions of hydrogen bonds. Both of the structures of ice grains in the homogeneous and partitioned square patterns (except for grain boundaries and point defects) satisfy the ice rule, i.e., every water molecule is hydrogen-bonded to its four nearest intralayer neighbors, with every water molecule being a double donor and a double acceptor of hydrogen bonds, as shown in Figure \ref{fig:fig02}. Most of the hydrogen bonds have switched to the in-plane configuration for the formation of the square ice structure. There are also some hydrogen atoms that lie between the two layers, which can form hydrogen bonds with oxygen atoms in adjacent layers and connect the two ice layers. The orientations of water molecules at the grain boundaries between the two regions are random, indicating that the partitioned square ice is a more unstable crystalline state with higher energy than the homogeneous square ice.

We also performed another series of MD simulations to consider the possibility of conversion from partitioned structure to homogenous square patterns by changing the pressure and the temperature. The lateral pressure $P$ was increased with a step of 0.1 GPa in the pressurization process, and the temperature $T$ was increased with a step of the 5 K in the heating process, respectively. The variation of the potential energy of the confined water during the pressurization and heating process is shown in Figure \ref{fig:fig03}(a) and (b), respectively. (We also performed simulations with different pressurization rates and heating rates, which show similar results to that in Figure \ref{fig:fig03}(a) and (b). See Figure S3 in Supplementary Material for details.) As the lateral pressure increases, water can transform from the liquid phase to the square ice structure when the pressure exceeds 1.0 GPa, and the square ice will decompose into amorphous phase if the pressure is approximately above 4.0 GPa, as shown in Figure \ref{fig:fig03}(a). As the temperature increases, the partitioned SI structure will transform into the homogeneous SI structure at around 320 K, and the partitioned SI structure will melt into liquid phase eventually when the temperature exceeds 420 K, as shown in Figure \ref{fig:fig03}(b). In some cases, the transformation from the partitioned SI structure to the homogeneous SI structure occurs in the pressurization process, as shown in Figure S2 and Movie S1 in Supplementary Material.

Hysteresis simulations are also performed through the pressurization/depressurization process by increasing the lateral pressure until 2.0 GPa and then decreasing the lateral pressure back to 1 bar ($10^{-4}$ GPa), as well as through the heating/cooling process by increasing the temperature until 550 K and then decreasing the temperature back to 300 K. The pressurization/depressurization process is similar to the cooling/heating process of the confined water, and both exhibit a large hysteresis loop, as shown in Figure \ref{fig:fig03}(c) and (d), implying that the formation of the square ice is a first-order phase transition. It is worth noting that the homogeneous SI structure does not transform into the partitioned SI structure during the cooling process.

\subsection{Difference in the Square Icing Patterns Between Graphene Layers and Graphite Layers}
\begin{figure}
  \centering
  \includegraphics[width=0.8\columnwidth]{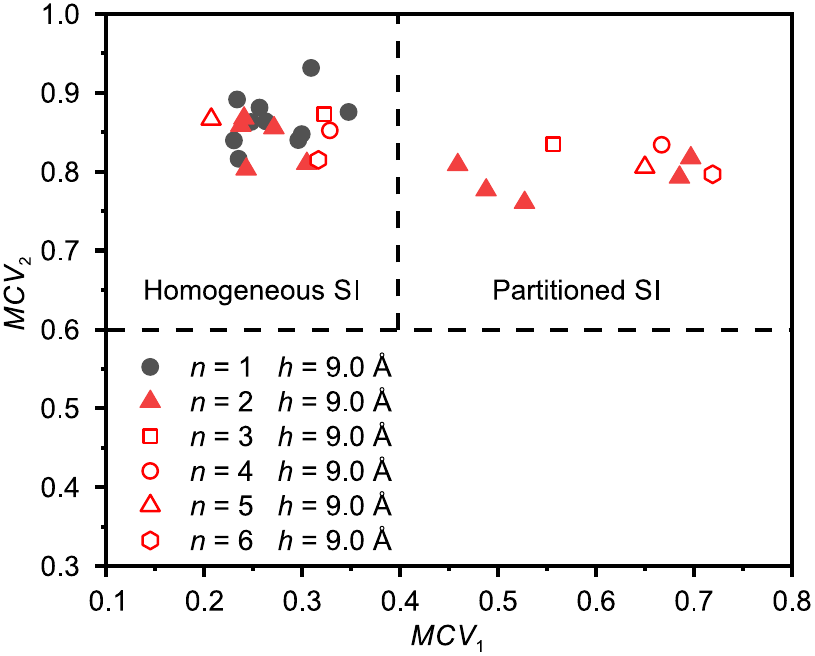}\\
  \caption{Square icing parameter map for different numbers of graphene monolayers ($n = 1 - 6$). The snapshots of the ten repeated simulations for $n = 2$ are shown in the Supplementary Material as Figure S3.}\label{fig:fig04}
\end{figure}

As discussed in the previous section, the number of graphene monolayers has a strong effect on the formation of partitioned square ice. To investigate the effect of the number of layers, we perform simulations with more graphene monolayers ($n = 3 - 6$), and the icing parametric diagram is shown in Figure \ref{fig:fig04}. Each case is repeated ten times by changing the initial velocity distributions of the water molecules while keeping the other settings identical (the snapshots of the ten repeated simulations for $n = 2$ are shown in Figure S3). By comparing the distributions of the icing parameters for $n = 1$ and $n = 2$, we find that it is energetically unfavorable for the water molecules confined between two parallel graphene sheets ($n = 1$) to form partitioned square icing patterns. In contrast, for the graphite ($n = 2$), the partitioned square icing patterns become common. The key factor is the different effects of graphite and graphene monolayer on water molecules, caused by the short-range nature of the interaction between water molecules and the carbon substrate. The diagram of the icing parameters in Figure \ref{fig:fig04} also shows that $n = 2$ does not always lead to the formation of the partitioned square icing patterns, and there is randomness in the formation of the partitioned structure.

From the distributions of icing parameters for $n$ = 2 -- 6 in Figure \ref{fig:fig04}, we can see that the partitioned square icing phenomenon is insensitive to the number of graphene monolayers as long as $n \geq 2$. This result can be explained by the short-range nature of the interaction between water molecules and the carbon substrate. In the force field that we use,\cite{Werder2008WaterCarbonInteraction} the water-carbon interaction is modeled by a 12-6 Lennard-Jones potential between the carbon and the oxygen atoms. The distance and the energy parameters are $\sigma = 0.328$ nm and $\epsilon = 0.114$ kcal/mol, respectively, and the distance between the graphene monolayers is 3.4 {\AA} in the graphite sheets. These force field parameters have been validated in many experiments and used in many MD simulations.\cite{Che2017SurfaceNanobubbles, Gordillo2000HydrogenBondStructure, Joly2011GiantLiquidSolidSlip, Werder2008WaterCarbonInteraction} According to these force fields, the second layer of carbon atoms exerts a strong force on the water molecules between the graphite sheets. However, from the third layer of carbon atoms, the force exerted on the water molecules will quickly vanish. Therefore, the short-range nature of the water-carbon interactions determines the strong difference between $n = 1$ and $n \geq 2$, as well as the insensitivity to $n$ when $n \geq 2$.

\subsection{Formation Process and Randomness Analysis}
\begin{figure}
  \centering
  \includegraphics[width=0.75\columnwidth]{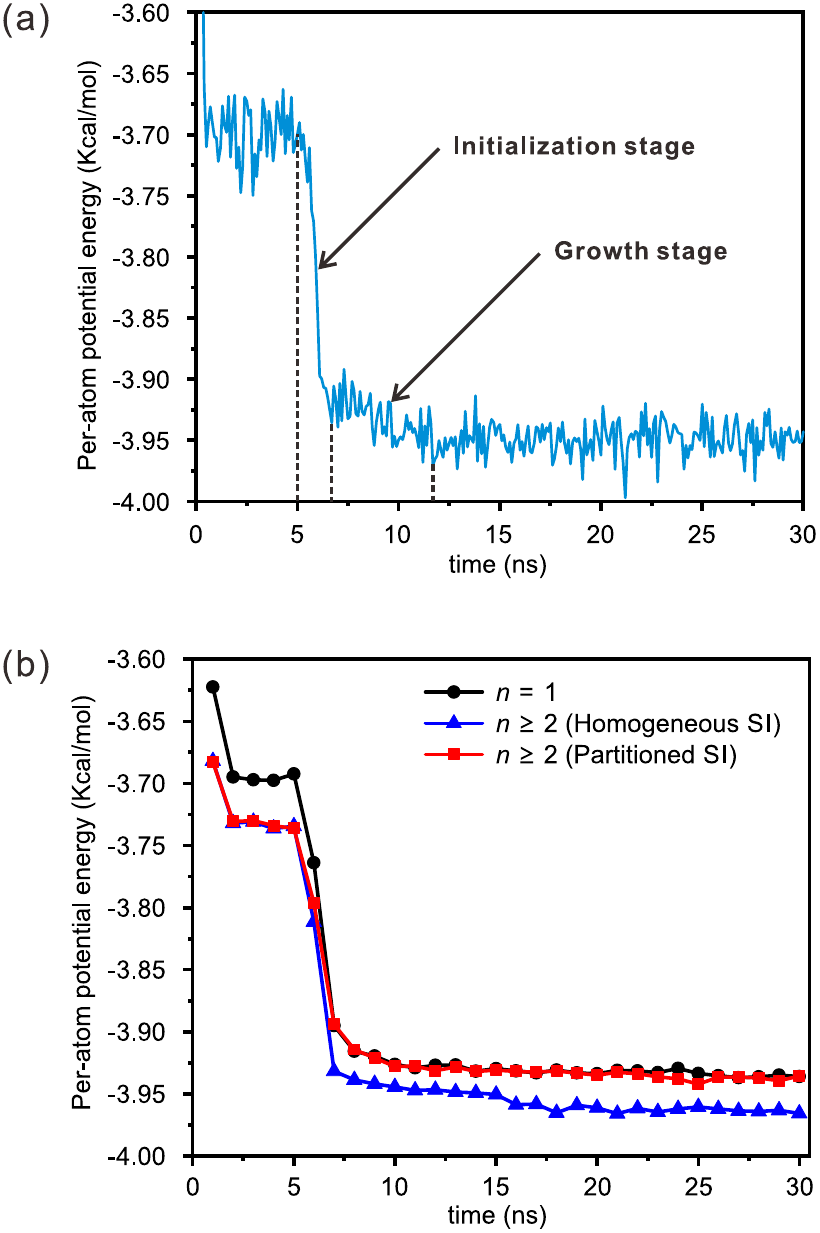}\\
  \caption{Variation of the potential energy of the confined water molecules. (a) Icing process for a typical case with $n = 2$. (b) Each curve represents the average potential energy of all cases with the same number of graphene monolayers and the same square icing patterns.}\label{fig:fig05}
\end{figure}

The variation of the potential energy of the confined water during the icing process for a typical case is shown in Figure \ref{fig:fig05}(a). After the first 5 ns, the lateral pressure $P$ is increased to 1.1 GPa and crystalline structures of water molecules begin to form. Even though the water molecules gain some extra energy from the high lateral pressure, more energy is released during the formation of hydrogen bonds, and the potential energy of water molecules drops rapidly until the system gradually stabilizes. The process of square ice formation lasts about $5 - 10$ ns in our simulation, including an initialization stage and a growth stage as shown in Figure \ref{fig:fig05}(a). In the initialization stage, water molecules change from a disordered liquid state to the locally ordered structure, while in the growth stage the locally ordered structure gradually expands until filling the whole space in the capillary. After that, the icing patterns remain stable eventually.

The variation in the potential energy shows different features for the formation of different ice structures, as shown in Figure \ref{fig:fig05}(b). All cases are taken into account in Figure \ref{fig:fig05}(b), and the cases with the same number of graphene monolayers and homogeneous/partitioned square icing patterns are grouped, and their potential energy is averaged per nanosecond. In the first 5 ns, the lateral pressure $P$ is kept at 1 bar ($10^{-4}$ GPa) for the equilibration. Water molecules confined between two parallel graphene sheets ($n = 1$) have significantly higher potential energy than those confined between graphite sheets ($n \geq 2$). According to the force field, the second layer of carbon atoms exerts a strong attractive force on the water molecules between the graphite sheets, which is different from the repulsion caused by the first layer of carbon atoms. As a result, water molecules gain higher potential energy from the carbon sheets when $n = 1$ at a fixed temperature. After the pressurization at 5 ns, water molecules gain extra energy from the high lateral pressure and, in the meantime, they release some energy to form hydrogen bonds, due to the formation of square ice. When the system gradually stabilizes after 20 ns, the water molecules for $n = 1$ and the water molecules with homogeneous square icing patterns for $ n \geq 2$ have almost the same difference in the potential energy compared to that before the pressurization. This is because they gain the same amount of energy from the same lateral pressure and consume the same amount of energy to form the same homogeneous square icing patterns in the pressurization process. Different from the situation for $n = 1$ in which water molecules have enough potential energy to support the formation of all hydrogen bonds, water molecules for $ n \geq 2$ have significantly lower potential energy, making it energetically unfavorable to form unidirectional, homogeneous icing patterns. Therefore, for $ n \geq 2$, the partitioned square icing structures with disordered local regions are more energetically favorable to form.

\begin{figure*}
  \centering
  \includegraphics[width=1.5\columnwidth]{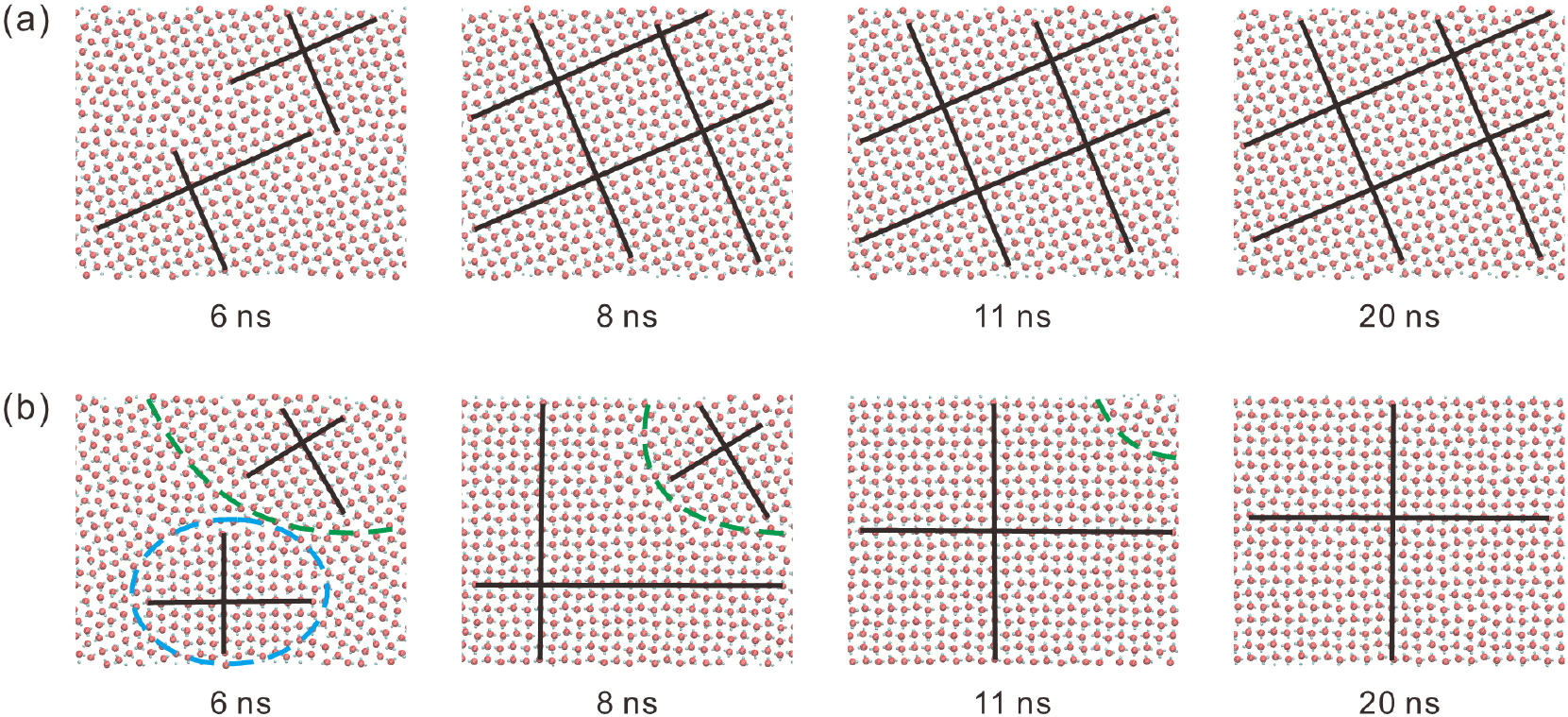}\\
  \caption{Two typical scenarios of square ice formation for $n = 1$. The black solid line is a guide to the eye of the icing direction, the green and the blue dashed curves are to indicate the local icing regions of different orientations. A video clip corresponding to Figure \ref{fig:fig06}(b) is available in the Supplementary Material as Movie S4.}\label{fig:fig06}
\end{figure*}

\begin{figure*}
  \centering
  \includegraphics[width=1.5\columnwidth]{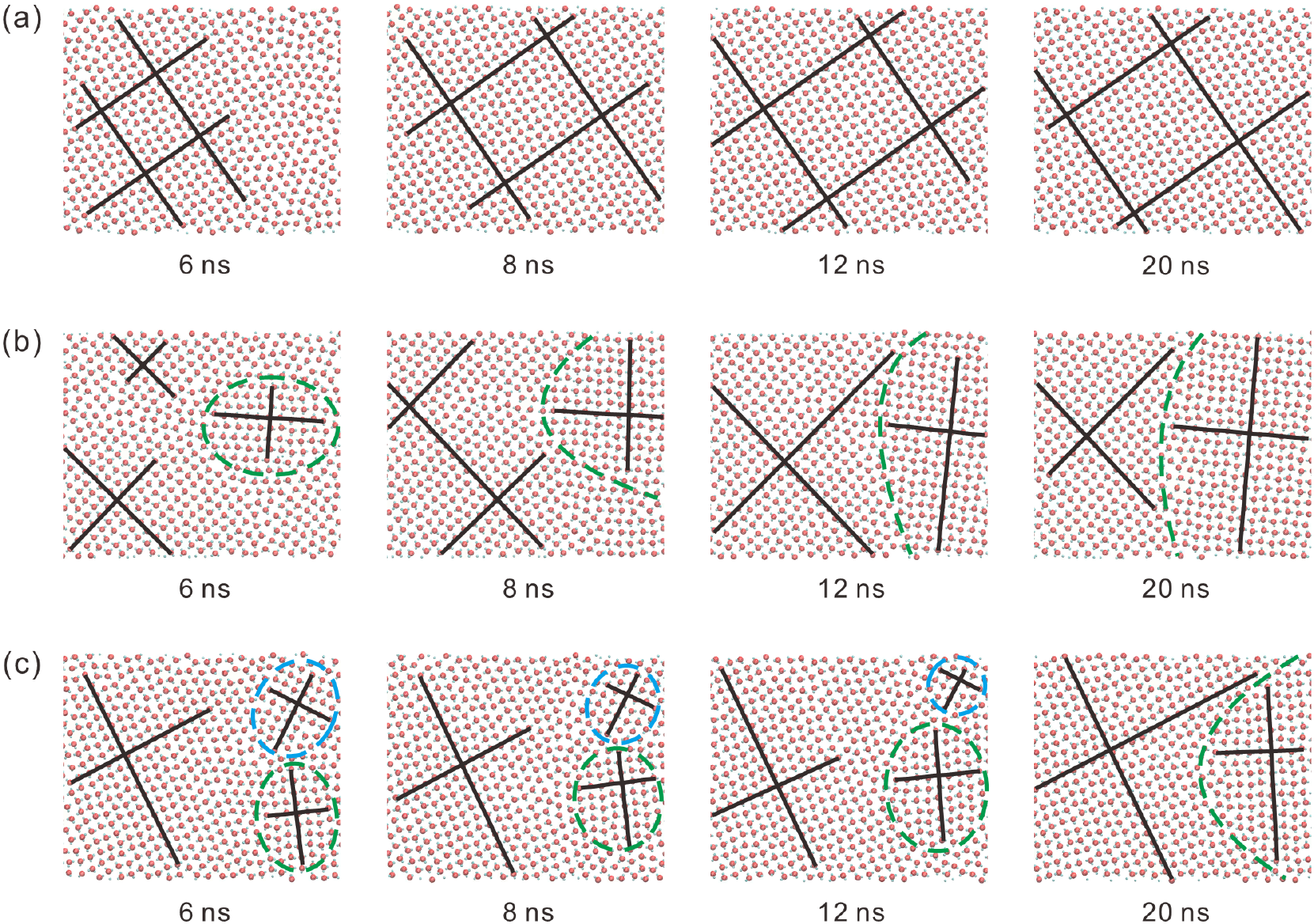}\\
  \caption{Three typical scenarios of square ice formation for $ n \geq 2$. The green and the blue dashed curves indicate the local icing regions with different orientations. Video clips corresponding to Figure \ref{fig:fig07}(b) and (c) are available in the Supplementary Material as Movies S5 and S6.}\label{fig:fig07}
\end{figure*}

For $n = 1$ (i.e., the water molecules are confined between two parallel graphene sheets), there are two typical scenarios of square ice formation, and their distributions of water molecules during the formation of square ice are shown in Figure \ref{fig:fig06} and Movie S4. In Figure \ref{fig:fig06}(a), square ice forms without partitioned patterns during the initialization stage. Therefore, in the subsequent growth stage, the square icing pattern is always homogeneous and eventually becomes stabilized. In contrast, in Figure \ref{fig:fig06}(b), there are two regions with different orientations of square icing patterns during the initialization stage. However, in the growth stage, one of them gradually becomes smaller and finally disappears, resulting in the homogeneous square icing pattern in the end. These typical processes reveal that for $n = 1$, partitioned square icing patterns will finally disappear during the growth stage because of the poor stability even if they are formed during the initialization stage. In the simulations, the energy gain of water molecules comes from graphene sheets and lateral pressure. The critical value of lateral pressure for ice formation can be estimated by the adhesion according to the previous study\cite{Algara2015SquareIce} ($P_W = E_W/d \approx 0.9$ GPa, where $E_W \approx 30$ meV/{\AA}$^{2}$ is the difference between graphene-graphene and graphene-water adhesion energies per unit area, and $d \approx 5.6$ {\AA} is the difference in the distance between plates due to the trapped water molecules), and confirmed by calculations of the per-atom potential energy (about 1.0 -- 1.1 GPa as shown in Figure S2 in Supplementary Material). Therefore, after the pressurization to 1.1 GPa for $n = 1$, water molecules have enough potential energy to support the formation of all hydrogen bonds, which can drive water molecules at the grain boundaries between the two regions to change from a disordered liquid state to the ordered crystalline state, and the homogeneous square icing patterns will eventually form.

For $ n \geq 2$ (i.e., the water molecules are confined between two parallel graphite sheets), there are also several typical scenarios of square ice formation, and their distributions of water molecules during the formation of square ice are shown in Figure \ref{fig:fig07} and Movies S5 \& S6. In Figure \ref{fig:fig07}(a), square ice forms without partitioned patterns during the initialization stage (same as the graphene case shown in Figure \ref{fig:fig06}(a)), leading to a homogeneous square icing pattern that becomes stable eventually. In contrast, in Figure \ref{fig:fig07}(b),here are some regions with square icing patterns in different orientations during the initialization stage. However, these small regions gradually enlarge and stabilize in the growth stage instead of becoming smaller and disappearing, finally leading to the partitioned square icing patterns. Even if more than two regions with different icing orientations are formed during the initialization stage in some particular cases, as shown in Figure \ref{fig:fig07}(c), a certain local region gradually enlarges and stabilizes while the other eventually disappears, forming the two-region partitioned icing patterns in the end. As analyzed in the previous section, water molecules have significantly lower potential energy when $n \geq 2$, which is not enough for all the water molecules at the grain boundaries to form the hydrogen bonds and change their disordered liquid state to the ordered crystalline state. Simulation results reveal that the randomness of the partitioned structure for $ n \geq 2$ originates from the initialization stage. Because of the short-range nature of the interaction between the water molecules and the carbon substrate, the effect of the carbon substrate determines the stability of the partitioned structure during the growth stage. The difference in stability eventually determines the probability of forming partitioned square icing patterns.

\begin{figure}
  \centering
  \includegraphics[width=0.85\columnwidth]{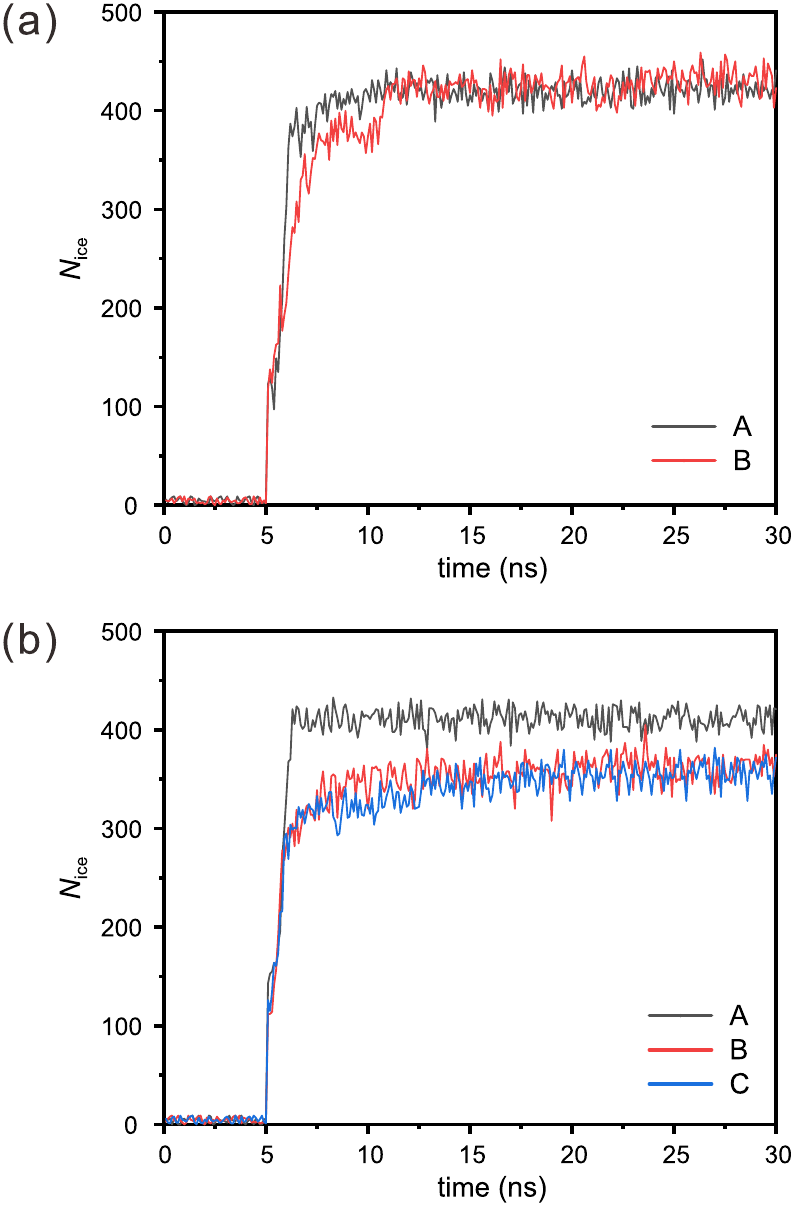}\\
  \caption{Number of ice water molecules as a function of time. (a) Two scenarios of square ice formation for $n = 1$, corresponding to Figure \ref{fig:fig06}(a) and (b), respectively. (b) Three scenarios of square ice formation for $n \geq 2$, corresponding to Figure \ref{fig:fig07}(a), (b), and (c), respectively.}\label{fig:fig08}
\end{figure}

The mechanisms of square ice formation generalized from the typical scenarios are further confirmed and quantified by analyzing the time evolution of the number of ice molecules, as shown in Figure \ref{fig:fig08}. After the stabilization, the total number of water molecules between the two parallel graphene-based sheets is approximately 500, most of which form the ordered structure, and the rest are grain boundaries or point defects. In Figure \ref{fig:fig08}(a), the number of ice molecules for case A rises suddenly after pressurization (at 5 ns) and then remains almost constant, corresponding to that square icing pattern is always homogeneous and eventually becomes stabilized (Figure \ref{fig:fig06}(a)). The number of ice molecules for case B is significantly lower than case A after pressurization (at 5 ns), which means there are multiple regions with different orientations of square icing patterns and water molecules at the grain boundaries are disordered during the initialization stage. Then it rapidly increases until about 10 ns, corresponding to that, regions with different orientations become a homogeneous square icing pattern in the growth stage (Figure \ref{fig:fig06}(b)). Similarly, in Figure \ref{fig:fig08}(b), the numbers of ice molecules for cases A, B, and C correspond to the typical scenarios shown in Figure \ref{fig:fig07}(a), (b), and (c), respectively. The numbers of ice molecules for cases B and C are always lower than that of case A because the orientations of water molecules at the grain boundaries between the two regions are random, indicating that the partitioned structure is stable during the growth stage for $ n \geq 2$. The results show that carbon substrate of graphene and graphite determined the stability of the partitioned structure during the growth stage.

\subsection{Interaction Between Layers of Water Molecules}
Besides the force by the carbon sheets on the water molecules, the confinement dimension is another important factor for the formation of the square icing patterns, as shown in Figure S6 in the Supplementary Material. In all of the above simulations, the height of the graphene capillary $h$ (the separation between the two sheets in $y$ direction) is fixed at 9.0 {\AA}, which is appropriate for the formation of the bilayer square ice. To consider the effect of the confinement dimension, we reduced $h$ to 6.5 {\AA}, which is appropriate for the formation of the flat monolayer square ice. We also varied the number of graphene monolayers $n$ to find out the possibility of partitioned square icing patterns. However, we did not observe the partitioned square icing patterns for the single layer of water molecules. To further verify the numerical simulation, we repeated the simulation for another 9 cases by using different initial velocity distributions, and still did not obtained the partitioned square icing patterns.

\begin{figure}
  \centering
  \includegraphics[width=0.85\columnwidth]{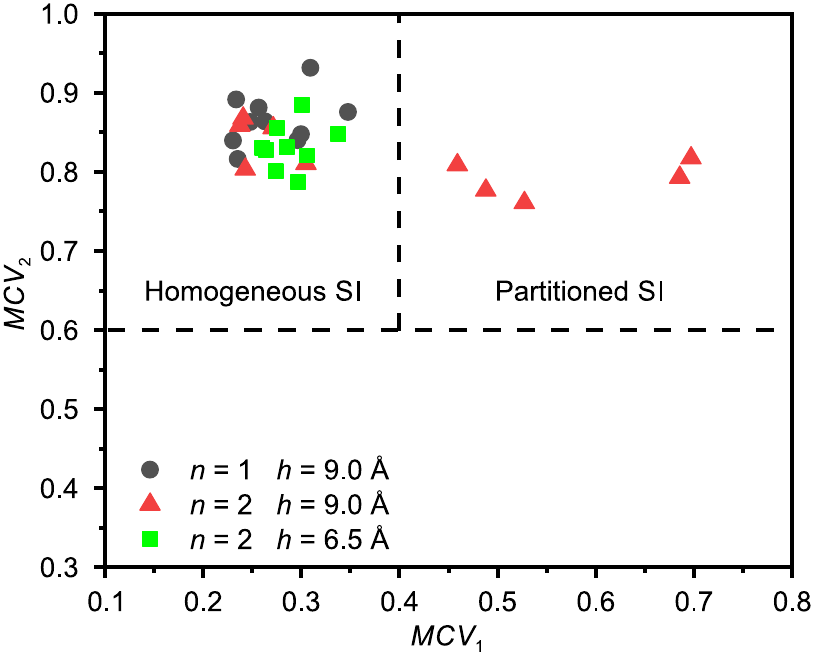}\\
  \caption{Square icing parameter map of different separation distances between two sheets ($h = 6.5 $ {\AA} and 9.0 {\AA}). The black solid circles represent the simulation results for $n = 1$ and $h = 9.0 $ {\AA}, the red solid triangles represent the simulation results for $n = 2$ and $h = 9.0 $ {\AA}, while the green solid squares represent the simulation results for $n = 2$ and $h = 6.5 $ {\AA}.}\label{fig:fig09}
\end{figure}

A square icing parameter map is produced for these simulation results, as shown in Figure \ref{fig:fig09}. From the results of single-water-layer cases, we can see that it is energetically unfavorable for the single-layer water molecules confined between two parallel graphite sheets to form partitioned square icing patterns, indicating that the interaction between different layers of water molecules is a dominant factor in partitioned square icing. Further analysis of the results for $h = 9.0 $ {\AA} reveals that both the top and the bottom layers of water molecules have the same behavior during the crystallization process and form identical square icing patterns after stabilization. This feature indicates that there is strong interaction between the layers of the water molecules that maintains these aligned structures. Like the hydrogen bonds between water molecules in the same layer, the interaction between the top and the bottom layers will consume part of the energy. When there is only one layer of water molecules between graphite sheets, the energy consumed for the interaction between layers is not needed. As a consequence, the water molecules have more potential energy to support the formation of all hydrogen bonds in the graphene capillary, even if they are confined between two parallel graphite sheets ($n \geq 2$).

\section{CONCLUSIONS}
In summary, partitioned square icing patterns are discovered and investigated in this study by using MD simulations. The different effects between graphite and graphene monolayer and the influences of confinement dimension and temperature are the most important factors for the 2D partitioned square ice. Different from water confined between graphene sheets, the partitioned square icing patterns are formed instead of the homogeneous square icing pattern when the graphene sheets turn into graphite. Moreover, the partitioned square icing phenomenon is insensitive to the number of graphene monolayers $n$ as long as $ n\geq 2$. For water molecules confined between graphite sheets ($ n\geq 2$), the partitioned square icing patterns do not always appear. The randomness comes from the initialization stage, and the carbon substrate of graphene and graphite determined the stability of the partitioned structure during the growth stage, because of the short-range nature of the interaction between water molecules and the carbon substrate. When the distance between the two graphite sheets decreases to 6.5 {\AA} to accommodate only one layer of water molecules, it is energetically unfavorable for the single layer of confined water molecules to form partitioned square icing patterns. This result indicates that the interaction between layers of water molecules is a dominant factor in partitioned square icing. The conversion from partitioned structure to homogenous square patterns is investigated by changing the pressure and the temperature. The pressurization/depressurization process and the cooling/heating process of the confined water exhibit a large hysteresis loop, implying that the formation of the square ice is a first-order phase transition. Based on the comprehensive MD simulations under different conditions, this study not only provides physical insights into the formation mechanism of the partitioned square icing patterns, but also will be helpful for the practical application in nanotribology, nanofluidic, and nanomaterials.

\section*{SUPPLEMENTARY MATERIAL}
Videos of confined water during the pressurization/heating process (Movies S1 -- S3), videos of typical square ice formations (Movies S4 -- S6), and figures of potential energy dependence of the lateral pressure and more partitioned square icing patterns (PDF).

\begin{acknowledgments}
This work was financially supported by the National Natural Science Foundation of China (No.\ 51920105010 and 52176083).
\end{acknowledgments}

\section*{AUTHOR DECLARATIONS}
The authors declare no competing financial interest.

\section*{DATA AVAILABILITY}
The data that support the findings of this study are available from the corresponding author upon reasonable request.

\nocite{*}
\bibliography{squreIce}

\end{document}